\documentclass[letterpaper,twocolumn,amssymb,floatfix]{revtex4}
\usepackage{times}
\usepackage{natbib}
\bibpunct {(}{)}{;}{a}{,}{,\,}
\usepackage{amsmath}
\usepackage{epsfig}

\begin{document}

\title{Winner-take-all selection in a neural system with delayed feedback
}

\author{Sebastian F.~Brandt}
\affiliation{Department of Physics, Washington University in St.~Louis, MO 63130-4899, USA}
\author{Ralf Wessel}
\affiliation{Department of Physics, Washington University in St.~Louis, MO 63130-4899, USA}
\date{September 14, 2007}

\begin{abstract}
We consider the effects of temporal delay in a neural feedback system with excitation and inhibition. The topology of our model system reflects the anatomy of the avian isthmic circuitry, a feedback structure found in all \linebreak[4] classes of vertebrates. We show that the system is capable of performing a `winner-take-all' selection rule for certain combinations of excitatory and inhibitory feedback.  In particular, we show that when the time delays are sufficiently large a system with local inhibition and global excitation can function as a `winner-take-all' network and exhibit oscillatory dynamics. We demonstrate how the origin of the oscillations can be attributed to the finite delays through a linear stability analysis.
\end{abstract}

\maketitle

\section{Introduction}
\label{Introduction}
In order to identify and react to behaviorally relevant objects in their visual environment, animals must be able to rapidly locate the positions of these objects in visual space.  This ability to select and orient towards the most salient part in a visual scene that may be cluttered with other, for the animal's survival less relevant objects, has evolutionary significance, as it permits the organism to detect quickly possible prey, predators, and mates \citep{Itti}. In standard models of selective visual attention, the stimulus is encoded in a `saliency map' that topographically represents the conspicuity of the stimulus over the visual scene. The most salient location is then chosen by a `winner-take-all' (WTA) network, i.e., by a neurally implemented maximum detector \citep{Koch}. In neuronal network models, these WTA networks are often realized as networks with lateral inhibition \citep{Amari,Kaski,Coultrip}, global inhibition \citep{Liu}, or local excitation and long distance inhibition \citep{Standage}. After the most active location, i.e., the `winner', in the saliency map has been chosen, attention should not, however, continue to be focused onto it. One way of allowing attention to shift, is to transiently inhibit neurons in the saliency map that correspond to the currently attended location, a strategy known as `inhibition of return' \citep{Itti2}.

\begin{figure}[t]
  \centering
    \includegraphics[width=\columnwidth]{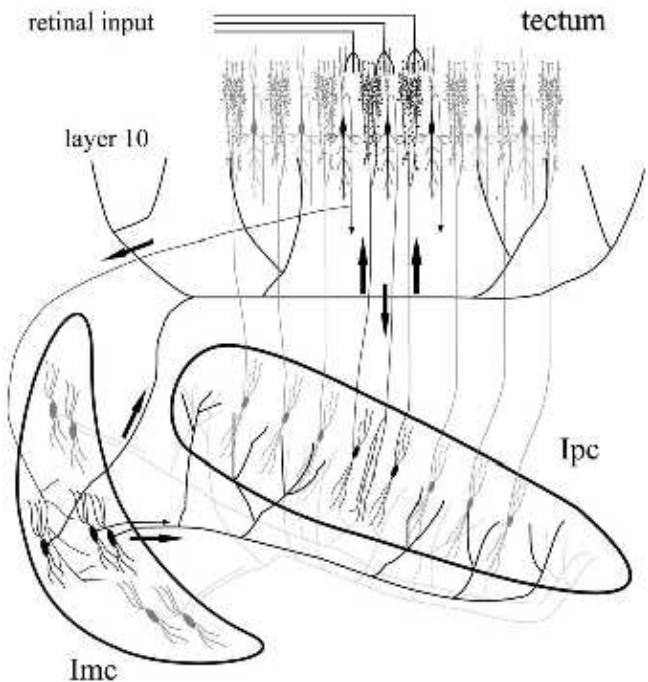}    
    \caption{Diagram of the isthmotectal feedback loop.  The Ipc is reciprocally connected with the TeO in a precise homotopic manner. Tectal neurons project topographically to the Ipc, and Ipc neurons project back to the corresponding tectal loci.  Imc receives a coarser topographic projection and projects back to the TeO and Ipc via widely ramifying terminal fields. Black represents visually activated neural elements. Reprinted with permission from \citet{Marin}.}
    \label{fig1}
\end{figure}

The homolog of the mammalian superior colliculus in non-mammalian vertebrates is the optic tectum (TeO). It is critically involved in localizing visual objects and in the  prep\-aration of orienting responses towards these objects \linebreak[4] \citep{Wurtz,Sparks}. In all classes of vertebrates, the TeO is reciprocally connected with the nucleus isthmi (NI), which is homologous to the parabigeminal nucleus \citep{Diamond} in mammals.  In the avian visual pathway, the NI consists of three subnuclei: the nucleus pars parvocellularis (Ipc), the nucleus pars magnocellularis (Imc), and the nucleus pars semilunaris (SLu) \citep{WangY1,Luksch}. In both Ipc and Imc the projection from the tectum is topographically organized such that the retinotopic map is preserved in both nuclei, with the projection to the Imc being somewhat coarser than for the Ipc \citep{WangY1}. In contrast, the isthmic projections back to the TeO are very different for Ipc and Imc. Ipc neurons project back to the TeO in a highly precise homotopic manner, i.e., the axons of each Ipc neuron terminate in that part of the optic tectum from which their visual inputs come \citep{Luksch}. Imc, on the other hand, has two populations of neurons, which both make heterotopic projections but only to the TeO or Ipc, respectively \citep{WangY1}. The three-nuclei circuitry consisting of TeO, Ipc, and Imc is shown in Fig.~\ref{fig1} \citep{Marin}.  Due to latencies arising from synaptic process and the spatial separation of the nuclei, the coupling between TeO and NI cannot be considered instantaneous.  Rather, finite temporal delays exist \citep{WangY1,Luksch}. Furthermore, delays can arise from the dynamical properties of the systems involved.  For instance, \citet{Andersen} report stimulus-dependent onset latency of recurrent inhibition in the cat hippocampus, and these findings were later explained by \citet{Hauptmann}.  It has been known for some time that temporal delays can cause an otherwise stable system to oscillate \citep{Heiden,Coleman,Hadeler} and may lead to bifurcation scenarios resulting in chaotic dynamics \citep{Wischert,Schanz}.  Therefore, finite delays are an essential property of any realistic model of a neuron population \citep{Milton}. 
\begin{figure}[t]
  \centering
    \includegraphics[width=\columnwidth]{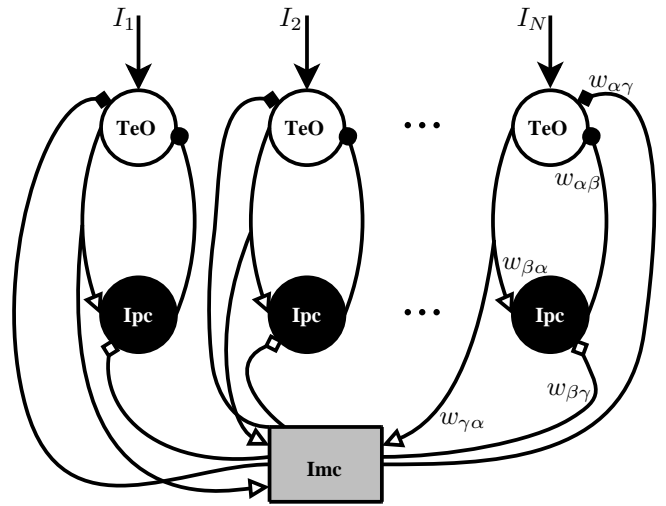}    
    \caption{Diagrammatic representation of our model for the isthmo\-tectal feedback loop. Neurons in the TeO and Ipc, which project topographically, are modeled as individual units. Due to their diffuse projections, Imc neurons are combined to form a feedback kernel (cf.~Tab.~\ref{tab1}).}
    \label{fig2}
\end{figure}

The synaptic effect of the recurrent projections from the Ipc and Imc cells onto their target cells is less well understood than their anatomical organization.  The available evidence suggests that Ipc neurons are cholinergic, whereas Imc neurons have been shown to express gamma-amino\-butyric acid (GABA) as their main neurotransmitter. Thus, according to the usual role of acetylcholine and GABA, one might speculate that Ipc and Imc neurons mediate excitation and inhibition onto their target cells, respectively. \citet{Marin} posit that ``the three-nuclei circuitry [...] may constitute
\begin{table}[b]
\caption{Components of the isthmotectal feedback loops and abbreviations.  We use the greek indices $\alpha$, $\beta$, and $\gamma$ to denote TeO, Ipc, and Imc, respectively.  Furthermore, neurons are numbered such that indices $1$ through $N$ refer to the TeO, $N+1$ through $2N$ refer to the Ipc, and the index $2N+1$ refers to the Imc.}
\centering
\label{tab1}
\begin{tabular}{llll}
\hline\noalign{\smallskip}
optic tectum & nucleus isthmi pars & nucleus isthmi pars\\ 
& parvocellularis &  magnocellularis  \\
\noalign{\smallskip}\hline\noalign{\smallskip}
TeO & Ipc & Imc \\
$\alpha$ & $\beta$ & $\gamma$ \\
$1\ldots N$ & $N+1 \ldots 2N$ & $2N+1$ \\
\noalign{\smallskip}\hline
\end{tabular}
\end{table}
a winner-take-all network \citep{Koch} in which local visual inputs to the Ipc are augmented by the re-entrant loop among tectal and Ipc neurons, combined with broad inhibition of the rest of the Ipc by Imc neurons.'' This argument seems immediately plausible, however, electrophysiological experiments \citep{WangSR1,WangYC,WangY2} suggest that the synaptic effects of Ipc and Imc are actually converse to this scenario and that the Ipc mediates inhibition whereas the Imc has an excitatory effect.  Given the anatomical organization of the recurrent projections from Ipc and Imc, it is not fully intuitive how the system could function as a WTA network when Imc is excitatory and Ipc inhibitory.  Nevertheless,\linebreak[4] \citet{WangSR2} considers this possibility: ``The positive and negative feedback loops formed between the tectum and NI may work together in a winner-take-all network, so that the positive feedback loop could provide a powerful augmentation of activated loci, while the negative feedback loop may strongly suppress the others [...]. For example, Imc could enhance the visual responses of tectal cells to target locations or stimulus features, while Ipc may suppress those to other locations or features in the visual field.'' The aim of this work is to investigate possible mechanisms for WTA selection in the isthmotectal feedback loop through a computational model. In this context, we do not refer to the term WTA in its most strict sense, which would imply that only the neuron with the strongest input exhibits a nonzero firing rate; rather, we speak of WTA behavior when the firing rates of neurons with weaker inputs are suppressed relative to those with stronger input.  

In Sect.~\ref{Model}, we introduce our model of the isthmic system, and we analyze its response dynamics for different temporal delays and different combinations of excitation and inhibition in Sect.~\ref{ResDyn}. In Sect.~\ref{Comp}, we compare the efficiency of WTA selection for these combinations. In Sect.~\ref{LSA}, we employ a linear stability analysis to show how the oscillatory dynamics that arise in the system can be attributed to the increasing delays. In Sect.~\ref{Sum}, we summarize our results.

\section{Model \label{Model}}
To explore the conjecture that the isthmotectal feedback loop functions as a WTA network, we consider a model system of coupled Hopfield neurons with temporal delays  \citep{Hopfield,Marcus}, as described by \citet{Ermentrout}. In this model, the temporal evolution of the membrane potential of the $i$th neuron (taken from rest potential), $V_i(t)$, is given by the first-order delay differential equation (DDE)
\begin{eqnarray}
\tau^{({\rm m})}_i \frac{d V_i(t)}{dt} = - V_i(t) + \sum_{j} w_{ij} r_j (t - \tau_{ij}) + I_i(t)\, .
\end{eqnarray}
Here, the membrane time constant for the $i$th neuron is denoted by $\tau^{({\rm m})}_i$, the synaptic connection weights for the projection from the $j$th to the $i$th neuron are $w_{ij}$, the temporal delay for this projection is $\tau_{ij}$, $r_j$ is the firing rate for the $j$th neuron and is linked to its voltage according to a nonlinear firing rate function, 
\begin{eqnarray}
r_j = S_j(V_j)\, ,
\end{eqnarray}
and $I_i(t)$ denotes an external input to the $i$th neuron.  To model the isthmic system, we assume that $N$ tectal neurons are reciprocally coupled to $N$ Ipc neurons and that the only neurons that receive external input are those in the TeO.  Furthermore, due to the broad and heterotopic nature of the projections from Imc, we combine the Imc neurons to a feedback kernel, which then projects globally to both TeO and Ipc. The topological structure of our model is depicted in Fig.~\ref{fig2}. 
To simplify our model, we make the following assumptions:  The synaptic weights for the projections TeO$\rightarrow$Ipc, \linebreak[4] TeO$\rightarrow$Imc, Ipc$\rightarrow$TeO, Imc$\rightarrow$TeO, Imc$\rightarrow$Ipc, are the same for all neurons in each of these groups, and we denote them by $w_{\beta\alpha}$, $w_{\gamma\alpha}$, $w_{\alpha\beta}$, $w_{\alpha\gamma}$, $w_{\beta\gamma}$, respectively; all membrane time constants are identical, $\tau^{({\rm m})}_i = \tau^{({\rm m})}$ for all $i$, and we rescale time such that $\tau^{({\rm m})} = 1$; all delays are identical,  $\tau_{ij} = \tau$; all firing rate functions are identical $S_j(V_j) = S(V_j)$.  
Furthermore, we number our neurons such that the indices $i = 1, \, 2, \, \ldots,\, N$ refer to tectal neurons, the indices $i = N+1,\, N+2,\, \ldots,\, 2N$ refer to Ipc neurons, and the index $i = 2N + 1$ refers to the Imc kernel (cf.~Tab.~\ref{tab1}).  Then, the dynamics of our system are described by the $2N + 1$ DDEs:
\begin{figure*}[t]
  \centering
    \includegraphics[width=0.8\textwidth]{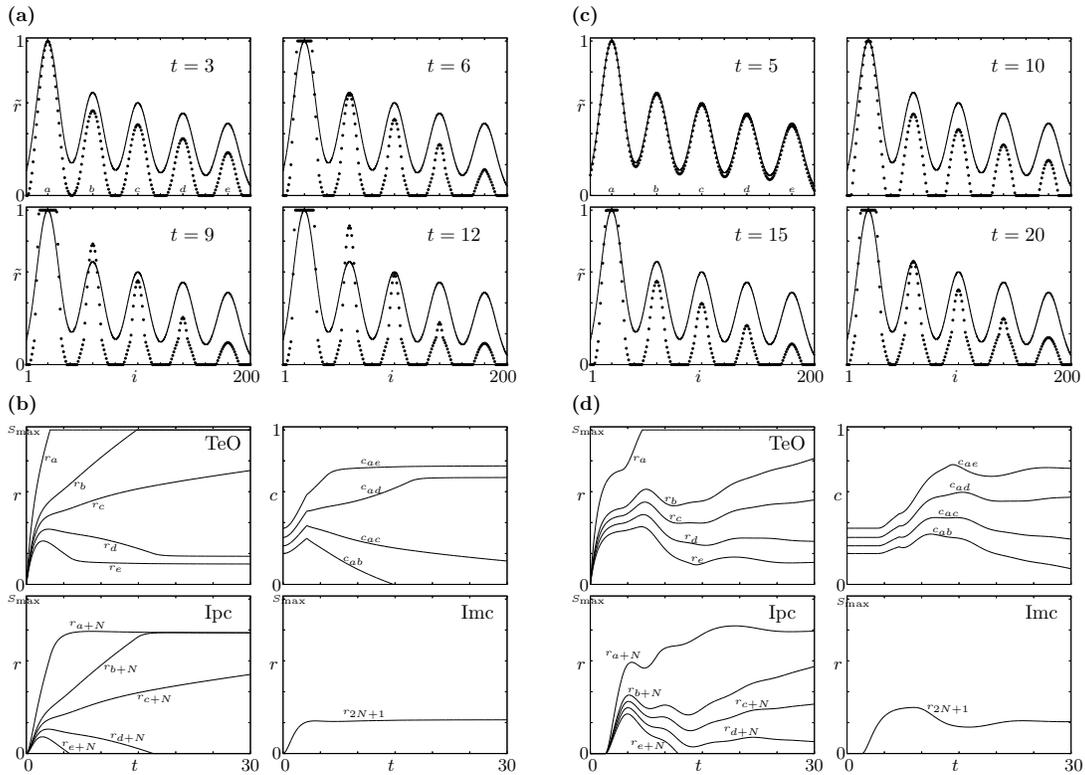}    
    \caption{WTA behavior and network dynamics for the case $(+,-,-)$, i.e., global inhibition and local excitation.  {\bf (a)} and {\bf (b)} are for the undelayed case, $\tau = 0$, while in {\bf (c)} and {\bf (d)} the delay is $\tau =2$. The dots in the snapshots in {\bf (a)} and {\bf (c)} show the normalized response of tectal cells, i.e., the quantity $\tilde{r}_i = r_i(t)/r_{\rm max}(t)$, where $r_{\rm max}(t)$ denotes the maximum of all tectal firing rates at time $t$, while the solid line shows the normalized input, i.e., the quantity $I_i/I_{\rm max}$. In both {\bf (b)} and {\bf (d)}, the first plot shows the firing rate dynamics of the neurons $a$ through $e$ in the TeO, the second plot depicts the temporal evolution of the response contrast of tectal neurons $b$ through $e$ when compared with neuron $a$, the third plot shows the activity of the Ipc neurons receiving input from the tectal neurons $a$ through $e$, and the fourth plot shows the firing rate of the Imc.} 
\label{fig3}
\end{figure*}
\begin{figure*}[t]
  \centering
    \includegraphics[width=0.8\textwidth]{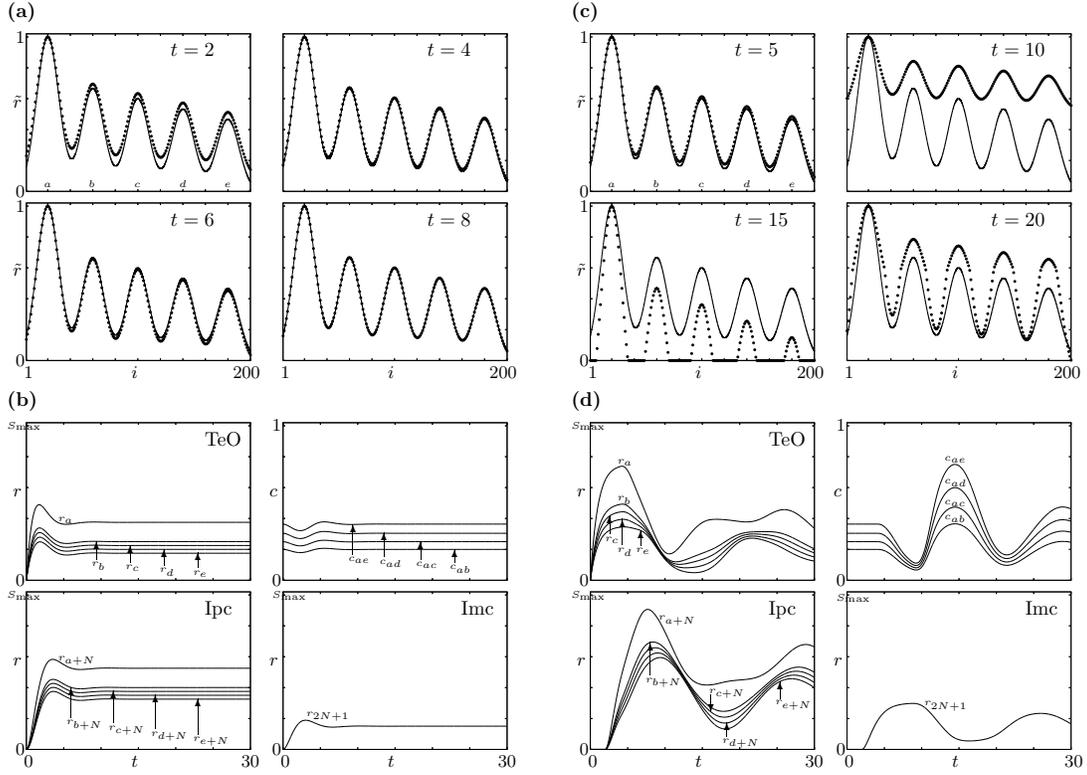}    
     \caption{WTA behavior and network dynamics for the case $(-,+,+)$, i.e., global excitation and local inhibition. Same as Fig.~\ref{fig3}, but for inversed signs of $w_{\alpha\beta}$, $w_{\alpha\gamma}$, and $w_{\beta\gamma}$.} 
\label{fig4}
\end{figure*}
\begin{eqnarray}
 \frac{d V_i(t)}{dt} &=& - V_i(t) + w_{\alpha\beta} r_{i+N} (t - \tau) \nonumber \\ && \hspace{-15mm} + w_{\alpha\gamma}r_{2N+1} (t - \tau) + I_i(t)\, , \quad i = 1, 2, \ldots, N  \,, \nonumber \\
 \frac{d V_i(t)}{dt} &=& - V_i(t) + w_{\beta\alpha} r_{i-N} (t - \tau) \nonumber \\ 
  && \hspace{-15mm} + w_{\beta\gamma}r_{2N+1} (t - \tau)\, ,  \quad \hspace{-2mm}i = N+1, N+2, \ldots, 2N, \nonumber \\
\frac{d V_{2N + 1}(t)}{dt} &=& - V_{2N + 1}(t) + w_{\gamma\alpha} \sum_{i = 1}^N  r_i(t - \tau) \,.  \label{DDEsys} 
\end{eqnarray}
For the firing rate function we choose the piecewise linear function,
\begin{equation}
r_j = S(V_j) = 
\begin{cases} 
0 &  \hspace{-2mm} \text{for $V_j<V_T$,} \\
a(V_j -V_T) & \hspace{-2mm} \text{for $V_T \leq V_j \leq V_T + S_{\rm max}/a$,}\\
S_{\rm max} &  \hspace{-2mm} \text{for $S_{\rm max}/a + V_T < V_j$.}
\end{cases}
\end{equation}
Finally, we make the simplifying assumptions $\left|w_{\beta\alpha}\right|=\linebreak[4] N \left|w_{\gamma\alpha}\right| =  \left|w_{\alpha\beta}\right| = \left|w_{\alpha\gamma}\right| = \left|w_{\beta\gamma}\right| = 1/ a$ and $V_T = 0$.  The signs of the synaptic weights determine whether a projection is excitatory or inhibitory and since the tectal cells mediate excitation, we have $w_{\beta \alpha}$, $w_{\gamma\alpha}$ $>0$. \citet{Marin} and \citet{WangSR2} both discuss scenarios in which WTA behavior arises from an interplay of excitation and inhibition in the isthmotectal feedback loop, and we are therefore interested in the cases where Ipc and Imc have adversary effects onto the TeO.  Thus, four cases remain to be studied, which can be characterized according to the signs of $(w_{\alpha\beta},\, w_{\alpha\gamma}, w_{\beta\gamma})$ as $(+,-,-)$, $(+,-,+)$, $(-,+,+)$, and $(-,+,-)$. In our model, the first two of these cases correspond to global inhibition and local excitation of tectal cells through feedback, whereas the latter two correspond to global excitation and local inhibition of the cells in the TeO. 

\section{Response dynamics \label{ResDyn}}
In the following, we investigate the dynamical response behavior of a network consisting of $N = 200$ (initially quiescent) neurons to a static stimulus. We choose an input consisting of five superimposed Gaussians with peaks at $i = 20,\, 60,\, 100,\, 140,$ and $180$, and peak values of $0.75$, $0.5$, $0.45$, $0.4$, and $0.35$, respectively.  The (normalized) stimulus is
\linebreak[4] 
shown, e.g., in Fig.~\ref{fig3} (a). In particular, we are interested in the firing rates of those neurons whose positions coincide with the peaks in the stimulus, and in order to abbreviate our notation we denote their indices as $a$, $b$, $c$, $d$, and $e$ according to descending strength of their respective inputs.

\subsection{Global inhibition, local excitation}
We first consider the cases $(+,-,-)$ and $(+,-,+)$.  In this situation, our network is similar to the circuit considered in \citep{Liu}. Therefore, we expect that it can perform a reasonably accurate WTA selection. Figure~\ref{fig3} shows the firing rate dynamics in response to the static input. The undelayed case is shown in Figs.~\ref{fig3} (a) and (b), whereas Figs.~\ref{fig3} (c) and (d) show the dynamics that result when $\tau = 2$.  To compare differences between firing rates of tectal neurons we consider the contrast measure
\begin{eqnarray}
c_{ij} = \frac{\left|r_i - r_j\right|}{r_i + r_j}\, .
\end{eqnarray}
In particular, we are interested in the contrasts $c_{ab}$, $c_{ac}$, $c_{ad}$, and $c_{ae}$, whose temporal dynamics are depicted in the second plots of Figs.~\ref{fig3} (b) and (d).  From Fig.~\ref{fig3} we see that in the case $(+,-,-)$ the weakest inputs are suppressed efficiently, while the neurons receiving the strongest input are driven towards maximum firing. Inputs of intermediate strength (e.g., the one received by neuron $b$) are not suppressed. Thus, in the configuration $(+,-,-)$ the system can perform a WTA selection, but not with very good `resolution'. Furthermore, by comparing Figs.~\ref{fig3} (a) and (b) with Figs.~\ref{fig3} (c) and (d), we see that the temporal delay in the system has only little effect on its efficiency as a WTA selector.  The main effect of the delay is that it causes the system to evolve on a longer time scale, i.e., the steady state is not reached as fast as in the undelayed case.

For the case $(+,-,+)$ it turns out that the inhibition in the system is insufficient to compensate for the positive feedback in the recurrent coupling between TeO and Ipc, and even neurons that receive only weak inputs are driven towards maximum firing. Thus, in the configuration $(+,-,+)$ our model system does not function as a WTA network.

\subsection{Global excitation, local inhibition}
Now we consider the cases $(-,+,+)$ and $(-,+,-)$, which correspond to global excitation of the TeO through the Imc and local inhibition from the Ipc.  They thus correspond to the scenario described by \citet{WangSR2} and it does not seem immediately intuitive how WTA behavior could result in this configuration.  Indeed, for the case $(-,+,+)$, when there are no delays no WTA selection occurs.  The response dynamics for this case are shown in Figs.~\ref{fig4} (a) and (b).  The contrast in the firing rate response of neurons receiving inputs of different strengths is nearly identical to the contrast of the respective inputs during all phases of the system's temporal evolution.  Thus, neither are weaker inputs suppressed nor are stronger inputs augmented.  When we introduce delay into the system, however, its behavior changes drastically.  Figs.~\ref{fig4} (c) and (d) show the response dynamics for the case $\tau = 2$.  Note that the neurons' firing rates, as well as the contrasts between responses now exhibit oscillatory behavior.  The system can perform a WTA selection with reasonable accuracy, but only transiently, i.e., only during certain phases of its temporal evolution.  As a matter of fact, a phase of best WTA selectivity is preceded and followed by phases where the response contrast is even lower than that of the input.

Since standard models of selective attention usually require that the most salient stimulus not be a permanent `winner', but rather that it be suppressed once attention has been directed to it, the dynamical evolution of the response is an important characteristic. Consequently, it may actually be a desirable feature of a WTA network to only determine the `winner' transiently.

The case $(-,+,-)$ leads to similar results as in the case $(+,-,+)$.  When Ipc neurons are inhibited by the Imc, they cannot provide sufficient negative feedback to the TeO in order to prevent tectal neurons from being saturated through the positive feedback between TeO and Imc. Thus, in the configuration $(-,+,-)$ our model system does not function as a WTA network.

\section{Comparison of WTA selectivity \label{Comp}}

\begin{figure}[t]
  \centering
    \includegraphics[width=8.1cm]{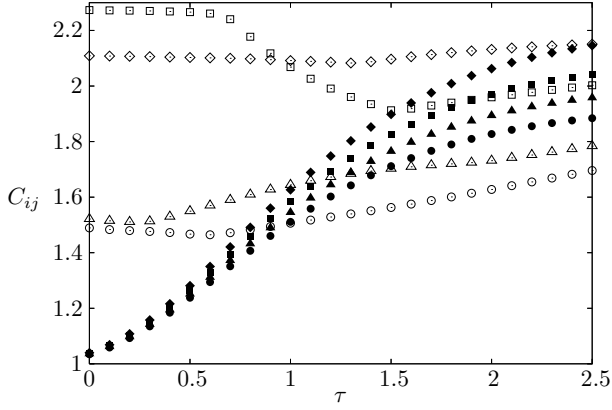}    
    \caption{Performance of the model system as a WTA network.  The maximal response contrast normalized to the input contrast is shown as a function of the delay $\tau$. Open symbols are for the case $(+,-,-)$, and filled symbols are for the case $(-,+,+)$. The contrast $C_{ij}$ is shown for $i = a$ and $j =b$ (circles), $j = c$ (triangles), $j = d$ (squares), and $j = e$ (diamonds).}
    \label{fig5}
\end{figure}

In order to quantify the performance of our model system as a WTA network, we consider the maximum in the response contrast between neurons $i$ and $j$, normalized to the contrast between the (constant) input $I$ to neurons $i$ and $j$ during the first $30$ membrane time constants of the system's temporal evolution:
\begin{eqnarray}
C_{ij} = \frac{I_i + I_j}{\left|I_i - I_j\right|} \max_{0\leq t \leq 30} c_{ij}\, \label{Cmax}.
\end{eqnarray}
This quantity is shown for the pairs $(a,b)$, $(a,c)$, $(a,d)$, and $(a,e)$ for the cases $(+,-,-)$ and $(-,+,+)$ and for different values of the time delay in Fig.~\ref{fig5}. As we expect from the results presented in Sec.~\ref{ResDyn}, in the case of global inhibition and local excitation, $(+,-,-)$, the system's performance as a WTA network, measured by the value of $C_{ij}$ depends only little on the time delay. Furthermore, we see that the system is efficient in suppressing weak inputs, whereas the response contrast for inputs of intermediate strength is less enhanced.  The performance for the case $(-,+,+)$, on the other hand, depends strongly on the temporal delay.  Furthermore, the ratio between the maximal response contrast and the input contrast is comparable for weak and intermediate inputs. When the delay is sufficiently large, the model system thus exhibits a better WTA `resolution' in the case of global excitation and local inhibition than for the inverse scenario, albeit only transiently.  

We have also investigated the role of parametric disorder in the system and find that it does not change our results qualitatively.  For instance, when the projection latencies are drawn randomly from a normal distribution with mean at 
$\tau=2$ and standard deviation of $0.2$ and the synaptic weights are disorderd using normal distributions with means at the default values and standard deviations of $10\%$ of these values, the maximum contrast as measured by $C_{ij}$ is comparable to the case without disorder.  Simulating ten different samplings of randomized delays and synaptic weights for the case $(-,+,+)$, we obtain the result $(C_{ab},\, C_{ac},\, C_{ad},\, C_{ae}) = (2.28 \pm 0.25,\, 2.50 \pm 0.30,\, 2.06 \pm 0.13,\, 2.34 \pm 0.12)$ (results are mean $\pm $ standarad error of the mean), which is to be compared with the values for the undisordered case \linebreak[4] $(C_{ab},\, C_{ac},\, C_{ad},\, C_{ae}) = (1.83,\, 1.89,\, 2.00,\, 2.06)$.  
   
\section{Linear Stability Analysis \label{LSA}}
\begin{figure}[t]
  \centering
    \includegraphics[width=\columnwidth]{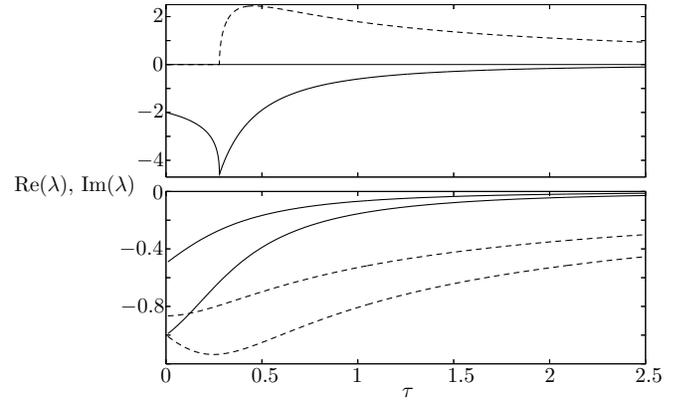}    
    \caption{Eigenvalues of $M$ as a function of the delay $\tau$. Real parts of eigenvalues are shown by solid lines, imaginary parts by dashed lines. The upper panel shows real and imaginary part of $\lambda_3$, and the lower panel shows real and imaginary of $\lambda_{1-}$ and $\lambda_{2-}$.}
    \label{fig6}
\end{figure}

We now aim to understand the origin of the delay-induced oscillatory dynamics in the case of global excitation and local inhibition through a stability analysis of the model system. To this end, we make the following ansatz, which, a posteriori, turns out to be correct.  We assume that for the chosen input the system of DDEs (\ref{DDEsys}) possesses a stationary point $V_i(t) = \bar{V}_i$ with $0 \leq \bar{V}_i \leq S_{\rm max}/a$ for $1 \leq i \leq 2N + 1$, and we can thus replace the voltages in the system (\ref{DDEsys}) according to the linear part of the firing rate function $S$.  The stationary point is then obtained by solving the equation
\begin{eqnarray}
{\bf \bar{V}} = a W {\bf \bar{V}} + {\bf I}\, . \label{lin}
\end{eqnarray}
Here, ${\bf \bar{V}}$ and ${\bf I}$ are $2N + 1$ column vectors (only the first $N$ entries of ${\bf I}$ are nonvanishing) and W is a $(2N + 1) \times (2N+1)$ matrix of the form
\begin{eqnarray}
W = \left( \begin{array}{ccc}
{\bf 0}_{N\times N} & w_{\alpha\beta} \openone_{N\times N} & w_{\alpha \gamma} \openone_{N \times 1} \\
w_{\beta \alpha} \openone_{N\times N} & {\bf 0}_{N\times N} & w_{\beta \gamma} \openone_{N \times 1} \\
 w_{\gamma \alpha} \openone_{1 \times N}& {\bf 0}_{1\times N} & 0 \end{array} \right) \, .
\end{eqnarray}
We find that for the case $(-,+,+)$, the matrix $\openone - a W$ is invertible and that the solution for the stationary point ${\bf \bar{V}} = (\openone - a W)^{-1}{\bf I}$ does indeed permit us to linearize the system (\ref{DDEsys}). In the case (+,-,-), however, it turns out that this solution yields values that lie outside of the linear regime of the firing rate function, and the linearization (\ref{lin}) is therefore not valid.  Next, we analyze the stability of the stationary point for the case $(-,+,+)$ by making the ansatz
${\bf V}(t) = {\bf \bar{V}} + {\bf c}e^{\lambda t}
$,
which leads to the equation $\left[a e ^{- \lambda \tau} W - (1 + \lambda) \openone \right] {\bf c} = {\bf 0}$. In order to determine the conditions for a nontrivial solution to this equation to exist, we must solve the characteristic equation for the matrix $M = a e^{-\lambda \tau} W - \openone$, i.e., we have to solve
\begin{eqnarray}
\det(M - \lambda \openone) = \left| \begin{array}{cc}
\tilde{M} & {\bf v} \\
{\bf u} & -(1+\lambda) \end{array} \right| = 0 \, , \label{CEq}
\end{eqnarray}
where we have introduced the abbreviations
\begin{eqnarray}
\tilde{M} &=& \left( \begin{array}{cc}
-(1 + \lambda)\openone_{N\times N} & a w_{\alpha \beta}e^{-\lambda \tau} \openone_{N\times N} \\
a w_{\beta \alpha}e^{-\lambda \tau} \openone_{N \times N} & -(1 + \lambda) \openone_{N\times N} \end{array} \right) \, , \\
{\bf u} &=& \left( a w_{\gamma \alpha} e^{-\lambda \tau}\openone_{1\times N}, {\bf 0}_{1\times N}  \right) \, , \\ 
{\bf v} &=& \left( \begin{array}{c}
a w_{\alpha \gamma} e^{-\lambda \tau} \openone_{N\times 1} \\
a w_{\beta \gamma} e^{-\lambda \tau} \openone_{N\times 1} \end{array} \right) \, . 
\end{eqnarray}
Solving (\ref{CEq}) is facilitated by applying the identity \citep{Pasolov}
$
\det(M) = - \det(\tilde{M})\left[ 1 + \lambda - {\bf u} \tilde{M}^{-1} {\bf v} \right]
$.
The inverse of $\tilde{M}$ is given by
\begin{eqnarray}
\tilde{M}^{-1} &=& - \frac{1}{(1 + \lambda)^2 - a^2 w_{\alpha \beta} w_{\beta \alpha} e^{-2 \lambda \tau}} \\  
&& \times \left( \begin{array}{cc}
(1 + \lambda)\openone_{N\times N} & a w_{\alpha \beta}e^{-\lambda \tau} \openone_{N\times N} \\
a w_{\beta \alpha} e^{-\lambda \tau} \openone_{N \times N} & (1 + \lambda) \openone_{N\times N} \end{array} \right) \, , \nonumber
\end{eqnarray}
and we thus obtain 
\begin{eqnarray}
\det(M - \lambda \openone) &=& \left[a w_{\alpha \beta} w_{\beta \gamma} e^{- \lambda \tau} + w_{\alpha \gamma}(1+\lambda)\right]  \nonumber \\ 
&& \hspace{-18mm}\times\hspace{1mm}  Na^2w_{\gamma \alpha}e^{-2\lambda \tau}\left[(1+\lambda)^2 - a^2 w_{\alpha \beta} w_{\beta \alpha} e^{-2 \lambda \tau}\right]^{N-1} \nonumber \\ && \hspace{-18mm} - \hspace{1mm} (1+\lambda)\left[(1+\lambda)^2 - a^2 w_{\alpha \beta}w_{\beta \alpha}e^{-2 \lambda \tau}\right]^N \, .
\end{eqnarray}
For the case $(-,+,+)$ the characteristic equation simplifies to
$
\left[1 + e^{3 \lambda \tau}(1 + \lambda)^3\right]\left[e^{-2 \lambda \tau} + (1 + \lambda)^2\right]^N = 0
$.
Its solutions, the eigenvalues of $M$, are given by
\begin{eqnarray}
\lambda_{1 \pm} &=& - 1 + \frac{1}{\tau}W(\pm i \tau e^\tau)\, , \\
\lambda_{2 \pm} &=& - 1 + \frac{1}{\tau}W\left[\left(\frac{1}{2} \pm i\frac{\sqrt{3}}{2} \right)\tau e^\tau \right]\, ,  \\
\lambda_3 &=& -1 + \frac{1}{\tau}W(-e^\tau \tau)\,.
\end{eqnarray}
Here, $W(z)$ is the inverse function to $W^{-1}(z) = z e^z$, which is usually called the Lambert $W$-Function. Figure~\ref{fig6} shows the real and imaginary parts of the eigenvalues of $M$ as a function of the delay $\tau$. For all values of $\tau$ there is no eigenvalue with a positive real part and the fixed point ${\bf \bar{V}}$ is thus stable for arbitrary delays.  However, the real parts of the eigenvalues tend to zero faster than their respective imaginary parts as the delay is increased. Therefore, with increasing delay, the relaxation time for the system's return to the stationary point grows more rapidly than the time scale for oscillations. Thus, with increasing delay, the system will spiral toward the fixed point, explaining the observed oscillatory behavior. 

\section{Summary and discussion\label{Sum}}
We have investigated the circumstances under which the isthmic system 
can function as a WTA network.  We have constructed a rate-model
of the isthmotectal feedback loop and have analyzed the temporal evolution of 
the model system in response to a static stimulus.  We have shown that time delays can be crucial to the dynamical behavior of the system.  In particular, delay-induced oscillations can lead to transient WTA selection 
in our model. Finally, we have performed a linear stability analysis 
explaining the origin of the oscillatory behavior.

It has been conjectured for a long time that the isthmotectal feedback loop constitutes a WTA network.  Our results show that the isthmic circuitry is indeed set up to perform such a selection rule.  In the case where global inhibition and local excitation are present in the system, this result is quite intuitive.  However, a network with global excitation and local inhibition might not appear to be well-suited as WTA selector. Yet, precisely such a scenario was discussed in the literature.  Temporal delays can be crucial for the behavior of a dynamical system, and, as we have shown in our investigation, they are particularly important for the case of global excitation and local inhibition, as they induce transient WTA behavior in the network.  Transmission and synaptic delays for the projections between Ipc and TeO are estimated to be around 15ms \citep{Netzel}, whereas membrane time constants in the Ipc may be as short as a few milliseconds (J. Shao, personal communication), which is within the range of typical neuronal membrane time constants \citep{KochB}.  When the synaptic and transmission delays are of the same order of magnitude as the membrane time constants involved, the degree to which our model for the isthmic system functions as a WTA network, can depend crucially on the delays (cf.~Fig.~\ref{fig5}). Therefore, temporal delays should not be neglected when the neuronal dynamics of the isthmotectal feedback loop are assessed and its potential for WTA selection is discussed.

\section*{Acknowledgements}
We thank Charles Anderson and John Clark for fruitful discussions and critical reading of the manuscript.
This work was supported by NIH-EY 15678.


\begin{thebibliography}{99}

\bibitem[Amari and Arbib(1977)]{Amari}
Amari A, Arbib MA (1977) Competition and cooperation in neural nets. 
In: Metzler J (ed) {\it Systems Neuroscience}, Academic Press, New York, NY, pp 119--165.
%
\bibitem[Andersen et al.(1964)]{Andersen}
Andersen P, Eccles JC, L\o yning Y (1964) Location of postsynaptic inhibitory synapses on hippocampal pyramids. J.\ Neurophysiol.\ {\bf 27}: 608--610.
%
\bibitem[Coleman and Renninger(1976)]{Coleman}
Coleman BD, Renninger GH (1976) Periodic solutions of certain nonlinear integral equations with a time lag.
SIAM J.\ Appl.\ Math.\ {\bf 31}: 111--120. 
%
\bibitem[Coultrip et al.(1992)]{Coultrip}
Coultrip R, Granger R, Lynch G (1992) A cortical model of winner-take-all competition via lateral inhibition.
Neural Networks {\bf 5}: 47--54.
%
\bibitem[Diamond et al.(1992)]{Diamond}
Diamond IT, Fitzpatrick D, Conley M (1992) A projection from the parabigeminal nucleus to the pulvinar nucleus in {\it Galago}.
J.\ Comp.\ Neurol.\ {\bf 316}: 375--382.
%
\bibitem[Ermentrout(1998)]{Ermentrout}
Ermentrout GB (1998), Neural networks as spatio-temporal pattern-forming
systems.
Rep.\ Prog.\ Phys.\ {\bf 61}: 353--430.
%
\bibitem[Hadeler and Tomiuk(1977)]{Hadeler}
Hadeler KP, Tomiuk J (1977)
Periodic solutions of difference differential equations.
Arch.\ Rat.\ Mech.\ Anal.\ {\bf 65}: 87--95.
%
\bibitem[Hauptmann and Mackey(2003)]{Hauptmann}
Hauptmann C, Mackey MC (2003) Stimulus-dependent onset latency of inhibitory recurrent activity.
Biol.\ Cybern.\ {\bf 88}: 459--467.
%
\bibitem[an der Heiden(1979)]{Heiden}
an der Heiden U (1979) Delays in physiological systems.
J.\ Math.\ Bio.\ {\bf 8}: 345--364.
%
\bibitem[Hopfield(1984)]{Hopfield}
Hopfield JJ (1984) Neurons with Graded Response Have Collective Computational Properties like Those of Two-State Neurons.
Proc.\ Natl.\ Acad.\ Sci.\ U.S.A.\ {\bf 81}: 3088--3092.
%
\bibitem[Indiveri and Delbr\"uck(2002)]{Liu}
Indiveri C, Delbr\"uck T (2002), 
in: Liu S-C et al., {\it Analog VLSI: Circuits and Principles}, The MIT Press, Cambridge, MA, chap.~6.
%
\bibitem[Itti and Koch(2001)]{Itti}
Itti L, Koch C (2001) Computational modelling of visual attention.
Nat.\ Rev.\ Neurosci.\ {\bf 2}: 194--204.
%
\bibitem[Itti et al.(1998)]{Itti2}
Itti L, Koch C, Niebur EA (1998) A Model of Saliency-Based Visual Attention for Rapid Scene Analysis. 
IEEE T.\ Pattern.\ Anal.\ {\bf 20}: 1254--1259.
%
\bibitem[Kaski and Kohonen(1994)]{Kaski}
Kaski S, Kohonen T (1994) Winner-take-all networks for physiological models of competitive learning.
Neural Networks {\bf 7}: 973--984.
%
\bibitem[Koch(1999)]{KochB}
Koch C (1999) {\it Biophysics of computation: information processing in single neurons}, Oxford University Press, New York, NY.
%
\bibitem[Koch and Ullman(1985)]{Koch}
Koch C, Ullman S (1985) Shifts in selective visual attention: towards the underlying neural circuitry.
Hum.\ Neurobiol.\ {\bf 4}: 219--227.
%
\bibitem[Marcus and Westervelt(1989)]{Marcus}
Marcus CM, Westervelt RM (1989) Stability of analog neural networks with delay.
Phys.\ Rev.\ A {\bf 39}: 347--359.
%
\bibitem[Mar\'{i}n et al.(2005)]{Marin}
Mar\'{i}n G, Mpodozis J, Sentis E, Ossand\'{o}n T, Letelier JC (2005)
Oscillatory Bursts in the Optic Tectum of Birds Represent Re-Entrant Signals from the Nucleus Isthmi Pars Parvocellularis.
J.~Neurosci.~{\bf 25}: 7081--7089.
%
\bibitem[Milton(1996)]{Milton}
Milton J (1996) {\it Dynamics of Small Neural Populations}, Amer.\ Math.\ Soc., Providence, RI.
%
\bibitem[Netzel et al.(2006)]{Netzel}
Netzel U, Wessel R, Luksch H (2006) A midbrain feedback loop in a slice preparation:  electrophysiology of relevant elements. 36th annual SfN meeting, Atlanta, GA, abstract 240.3.  
%
\bibitem[Pasolov(1994)]{Pasolov}
Pasolov VV (1994) {\it Problems and Theorems in Linear Algebra}, American Mathematical Society, Providence, RI.
%
\bibitem[Schanz and Pelster(2003)]{Schanz}
Schanz M, Pelster A (2003) Analytical and numerical investigations of the phase-locked loop with time delay. 
Phys.\ Rev.\ E {\bf 67}: 056205-1--8.
%
\bibitem[Sparks(1986)]{Sparks}
Sparks DL (1986) Translation of sensory signals into commands for control of saccadic eye movements: role of primate superior colliculus.
Physiol.\ Rev.\ {\bf 66}: 118--171.
%
\bibitem[Standage et al.(2005)]{Standage}
Standage DI, Trappenberg TP, Klein RM (2005) Modelling divided visual attention with a winner-take-all network.
Neural Networks {\bf 18}: 620--627.
%
\bibitem[Wang(2003)]{WangSR2}
Wang S-R (2003) The nucleus isthmi and dual modulation of the receptive field of tectal neurons in non-mammals.
Brain.\ Res.\ Rev.\ {\bf 41}: 13--25.
%
\bibitem[Wang et al.(1995)]{WangSR1}
Wang S-R, Wang Y-C, Frost BJ (1995) Magnocellular and parvocellular divisions of pigeon nucleus isthmi differentially modulate visual responses in the tectum.
Exp.\ Brain Res.\ {\bf 104}: 376--384.
%
\bibitem[Wang et al.(2006)]{Luksch}
Wang Y, Luksch H, Brecha NC, Karten HJ (2006) Collumnar projections from the cholinergic nucleus isthmi to the optic tectum in chicks ({\it Gallus gallus}): A possible substrate for synchronizing tectal channels.
J.\ Comp.\ Neurol.\ {\bf 494}: 7--35.
%
\bibitem[Wang et al.(2004)]{WangY1}
Wang Y, Major DE, Karten HJ (2004) Morphology and connections of nucleus isthmi pars magnocellularis in chicks ({\it Gallus gallus}).
J.\ Comp.\ Neurol.\ {\bf 469}: 275--297.
%
\bibitem[Wang et al.(2000)]{WangY2}
Wang Y, Xiao J, Wang S-R (2000) Excitatory and inhibitory receptive fields of tectal cells are differentially modified by magnocellular and parvocellular divisions of the pigeon nucleus isthmi.
J.\ Comp.\ Physiol.\ A, {\bf 186}: 505--512.
%
\bibitem[Wang and Frost(1991)]{WangYC}
Wang Y-C, Frost BJ (1991)
Visual response characteristics of neurons in the nucleus isthmi magnocellularis and nucleus isthmi parvocellularis of pigeons.
Exp.\ Brain Res.\ {\bf 87}: 624--633.
%
\bibitem[Wischert et al.(1994)]{Wischert}
Wischert W, Wunderlin A, Pelster A, Olivier M, and Groslambert J (1994)
Delay-induced instabilities in nonlinear feedback systems.
Phys.\ Rev.\ E {\bf 49}: 203--219.
%
\bibitem[Wurtz and Albano(1980)]{Wurtz}
Wurtz RH, Albano JE (1980) Visual-motor function of the primate superior colliculus.
Annu.\ Rev.\ Neurosci.\ {\bf 3}: 189--226.
%
\end{thebibliography}
\end{document}